\documentstyle[preprint, aps]{revtex}
\begin{document}
\title{\bf Full capacitance matrix of coupled quantum dot arrays: static and 
dynamical effects}
\author{C. B. Whan, J. White, and T. P. Orlando} 
\address{Department of Electrical Engineering and Computer Science,\\
Massachusetts Institute of Technology, Cambridge, MA 02139}
\date{\today}
\maketitle
\begin{abstract}
We numerically calculated the full capacitance matrices for 
both one-dimensional (1D) and two-dimensional (2D) 
quantum-dot arrays. We found it is necessary to use the 
full capacitance matrix in modeling coupled quantum dot arrays due to 
weaker screening in these systems in comparison with arrays of 
normal metal tunnel junctions. The static soliton potential 
distributions in both 1D and 2D arrays are well approximated by 
the unscreened $(1/r)$ coulomb potential, instead of the exponential 
fall-off expected from the often used nearest neighbor approximation. 
The Coulomb potential approximation also provides a simple expression 
for the full inverse capacitance matrix of uniform quantum dot arrays. 
In terms of dynamics, we compare the current-voltage (I-V) 
characteristics of voltage biased 1D arrays using either the 
full capacitance matrix or its nearest neighbor approximation. 
The I-V curves show clear differences and the differences  
become more pronounced when larger arrays are considered. \\
\end{abstract}
\newpage
 
Much progress has been made in understanding the transport properties of 
single quantum dots\cite{kouwenhoven}. In contrast, much less is known 
about transport through multiple quantum dots coupled by tunnel barriers. 
Recent experiments have mostly concentrated on double-dot or triple-dot 
systems\cite{kamerink,sakamoto,waugh}, with the exception of 
Ref.~\cite{marcus} which is devoted to large two-dimensional (2D) 
quantum dot arrays. In analyzing quantum dot arrays, most 
authors\cite{kouwenhoven,marcus,middleton} assume that in the 
classical charging regime (i.e. $\Delta E < k_BT \ll E_C$, where $E_C$ 
is the charging energy of the dot and $\Delta E$ is the quantum energy 
level spacing), one can simply use the existing analysis for metal tunnel 
junction arrays\cite{bakhvalov-1d,bakhvalov-2d,uli,mooij}. The purpose 
of this letter is to explicitly demonstrate that even in the classical 
charging regime, an important distinction exists between quantum dot arrays 
in semiconductors and metal tunnel junction arrays, due to the 
difference in their abilities to screen electric charge.  

  In metal tunnel junction arrays\cite{delsing}, the junctions 
form very effective parallel plate capacitors due to 
the high dielectric constant of the barrier material (for example, 
in ${\rm Al/Al_{2}O_{3}/Al}$ junctions, $\epsilon \approx 10 \epsilon_0$ 
for the ${\rm Al_{2}O_{3}}$ barrier) as well as the extremely 
small barrier thickness (${\rm \sim 1 \; nm}$). As a consequence, 
the electrical potential due to an extra charge (the so-called soliton) 
placed on one of the metal islands in an array of tunnel junctions 
gets screened and decays rapidly with the distance away from the charged 
island. Until recently, most analyses\cite{bakhvalov-1d,bakhvalov-2d,uli,mooij} considered only the nearest neighbor junction capacitance $C$ and 
the self-capacitance of the junction electrodes $C_0$ when constructing 
the capacitance matrix of the array. For a one-dimensional (1D) array, 
this approximation leads to an exponential decay of the soliton 
potential away from the center, with a characteristic screening 
length\cite{bakhvalov-1d}, $\lambda = a \sqrt{C/C_0}$, where $a$ 
is the lattice constant of the array. 

For quantum dot arrays made by electrostatic confinement of 
two-dimensional electron gas (2DEG) in ${\rm GaAs/AlGaAs}$ 
heterostructures\cite{waugh}, the array is a co-planar structure 
in which all the dots reside in the 2DEG plane. The co-planar 
capacitors are far less effective than the parallel plates in 
terms of confining electric field. Therefore, in comparison with 
tunnel junction arrays, the field lines originating from one of 
the quantum dots are much less confined and can reach out to dots 
that are much further apart. A model that considers only the nearest 
neighbor capacitive coupling is unlikely to be accurate in this 
situation. We now give a more quantitative analysis of this problem.   

In our model, the quantum dots (small puddles of a 2DEG) are 
treated as thin circular shaped conducting plates, with diameter 
$D = 1 \; {\rm \mu m}$. The plates are arranged to form either 
1D or 2D arrays with lattice constant $a \equiv D + d = 1.1 \;\mu m$, 
where $d = 0.1 \; {\rm \mu m}$ is the closest separation between 
adjacent dots (or the tunnel barrier width). We believe these 
values are reasonable for arrays in the classical charging regime with 
weak inter-dot tunneling (i.e. the tunneling resistance, 
$R_T \gg R_Q \equiv h/e^2$). Once the array geometry 
is specified, we compute the full capacitance matrix ${\bf C}$ of the 
1D and 2D arrays using FASTCAP, an efficient capacitance extraction 
tool\cite{nabor}. 
  
The inverse of the capacitance matrix gives us the potential 
distribution in the array for a given charge distribution 
within the array. In particular, the potential distribution 
due to a soliton (antisoliton) located at dot $i$ in an $N \times 1$ 
array is given by,
\begin{equation}
\label{eq:pmat}
\left(
\begin{array}{c}
  \phi_{1} \\ \vdots \\ \phi_{i} \\ \vdots \\ \phi_{N}
\end{array}
\right) 
= {\bf P} \left(
\begin{array}{c}
  0 \\ \vdots \\ \pm e \\ \vdots \\ 0
\end{array}
\right),   
\end{equation}  
where, ${\bf P \equiv C^{-1}}$, is the inverse of the capacitance matrix.

In Fig.~\ref{fig:soliton}a, we show the potential distribution due 
to a soliton located at the center ($i = 11$) of a $21 \times 1$ 
series array, using both the full capacitance matrix and 
its nearest neighbor approximation. The nearest neighbor approximation is 
obtained by setting all the non-nearest-neighbor off-diagonal elements 
to zero in the full capacitance matrix. As we can see, the nearest neighbor 
approximation gives an exponentially decaying soliton potential as 
expected\cite{bakhvalov-1d}. However, the soliton potential distribution 
that we get from the full capacitance matrix decays much slower. 

As we pointed out earlier, due to the spread-out nature of quantum 
dot arrays, there is relatively weak screening of electrostatic 
potentials. Therefore we might expect the soliton potential 
distribution to follow the usual $1/r$ Coulomb potential at 
large distances. In Fig.~\ref{fig:soliton}b, we compare the 
full capacitance matrix soliton potential distribution with 
the Coulomb potential of a point charge $e$ located 
at the soliton position. We see that the soliton potential 
follows the simple $1/r$ law almost exactly in the entire range, 
except at the origin where the Coulomb potential is singular. 

In order to obtain an expression for the potential at the origin, 
$\phi_{i}(i)$, we recall that $\phi_{i}(i)$ is the electrical 
potential of dot $i$ when a unit charge is placed on it and all 
the other dots are left neutral. When a dot has charge $Q$ and is 
isolated in free space, its electrical potential will be, 
$\phi_{0} = Q/C_0$, where $C_0$ is the self-capacitance 
of an isolated dot. From simple dimensional analysis, 
we expect $C_0$ to have the form, $
C_0 = \alpha\pi\epsilon_{0}D$,  for a dot with diameter $D$. 
Using FASTCAP, we determine the numerical factor $\alpha \approx 1.23$. 
When other (neutral) conductors are brought nearby, it should not 
disturb the potential $\phi_0$ too much and therefore we take, 
$\phi_{i}(i) \approx e/C_0$. Thus, we have the following 
approximate expression for the soliton potential distribution 
in a 1D quantum dot array:
\begin{equation}
\label{eq:coulomb}
\phi_{j}(i) = \left\{ 
\begin{array}{ll}
\frac{e}{\alpha\pi\epsilon_{0}D}  & \mbox {\rm if} \; i = j;\\
\frac{e}{4\pi\epsilon_{0}a}\frac{1}{|i - j|} & \mbox {\rm if} \; i \ne j. 
\end{array}
\right.  
\end{equation}
The results from Eq.~(\ref{eq:coulomb}) are denoted by the cross ($+$) 
symbols in Fig.~\ref{fig:soliton}b.

Recently, Likharev and Matsuoka\cite{likharev-matsuoka} pointed out that 
even in tunnel junction arrays, the nearest neighbor approximation is 
questionable. By considering full capacitance matrices of tunnel junction 
arrays and comparing them with an analytical continuum model, they proposed
a phenomenological formula for the soliton potential distribution in 
1D tunnel junction arrays. We attempted to fit our results to their formula, 
the best fit gives essentially the same result as the simple Coulomb 
potential, except at the the origin their formula overestimates 
$\phi_{i}(i)$ by roughly a factor of 2. This is not surprising since 
their formula was obtained for tunnel junction arrays where $ \lambda \gg a$. In our quantum dot arrays, however, the opposite limit, $\lambda < a$, applies. 

We also carried out similar calculations for 2D arrays of quantum dots. 
In Fig.~\ref{fig:2d}, we plot the soliton potential distribution in a 
$11 \times 11$ array with the soliton located in the center, using both 
the full capacitance matrix and its nearest neighbor approximation. 
Again, we found that the soliton potential in the nearest neighbor 
approximation follows the expected form\cite{bakhvalov-2d,mooij},
$\phi(r) \propto e^{-r/\lambda}/\sqrt{r}$ for $\lambda < a$ (not shown 
in the figure). The 2D soliton potential, when full capacitance matrix is 
considered, is once again well-approximated by the simple formula 
\begin{equation}
\label{eq:coulomb-2d}
\phi(x,y) = \left\{ 
\begin{array}{ll}
\frac{e}{\alpha\pi\epsilon_{0}D}  & \mbox {\rm if} \; x = y = 0 \\
\frac{e}{4\pi\epsilon_{0}}\frac{1}{\sqrt{x^2 + y^2}} & \mbox {\rm otherwise}, 
\end{array}
\right.  
\end{equation}    
assuming the soliton is located at the origin of the coordinate system.

According to Eq.~(\ref{eq:pmat}), if we know the potential distributions 
for all possible soliton locations, we can determine the full inverse 
capacitance matrix. Therefore, for 1D and 2D uniform disk-shaped 
quantum dot arrays, we can approximately write down the inverse 
capacitance matrix, ${\bf P}$, without any numerical calculations 
[see Eqs.~(\ref{eq:coulomb}), and ~(\ref{eq:coulomb-2d})]:
\begin{equation}
\label{eq:Pmat}
{\bf P}_{ij} = \left\{ 
\begin{array}{ll}
\frac{1}{\alpha\pi\epsilon_{0}D}  & \mbox {\rm if} \; i = j;\\
\frac{1}{4\pi\epsilon_{0}}\frac{1}{|{\bf r}_i - {\bf r}_j|} & \mbox {\rm if} \; i \ne j. 
\end{array}
\right.  
\end{equation}
  
So far, our analysis assumes that our array is in free space (or in a 
uniform dielectric medium if we make the substitution 
$\epsilon_0 \rightarrow \epsilon$). In reality, 
the 2DEG plane is buried in a dielectric medium with 
$\epsilon \approx 13 \epsilon_0$, very close to ($\sim 300 \;{\rm\AA}$) the 
sample surface. Underneath the 2DEG, the same dielectric medium 
continues over the entire sample thickness (including the 
GaAs substrate, which is about 0.5 mm thick). Therefore in 
a more realistic model for a quantum dot array, we can treat the 
array as being located at the interface between the free space 
($\epsilon_0$) and an infinitely thick substrate with permittivity 
$\epsilon = 13 \epsilon_0$. We can calculate the capacitance matrix 
of this system using FASTCAP assisted by the static image method. 
The result amounts to making the simple substitution, 
$\epsilon_0 \rightarrow \epsilon/2$, in the above analysis.

Having shown that the the full capacitance matrix is needed to 
model the static soliton potential in quantum dot arrays, we now 
briefly address the effect of full capacitance matrix on dynamical 
properties. In particular, we compute the current-voltage (I-V) 
characteristics of voltage biased arrays using Monte Carlo 
method\cite{bakhvalov-1d,bakhvalov-2d,uli}. In Fig.~\ref{fig:iv}, 
we compare I-V curves for a $21 \times 1$ array computed using 
the full capacitance matrix and the nearest neighbor approximation. 
As we can see in Fig.~\ref{fig:iv}, the two I-V curves 
clearly show many differences. The threshold voltages are not exactly 
the same, and the fine structures are different and they become more 
pronounced if we consider larger arrays. At high voltage, the two curves 
merge and become nearly linear, as shown in the inset of 
Fig.~\ref{fig:iv}. The detailed analysis of the I-V curves, 
as well as some of the relevant numerical techniques, will be the subject 
of a future publication. Nevertheless, we see that the full capacitance 
matrix is necessary in the dynamical simulation of these arrays.

We acknowledge fruitful discussions with David Carter and 
Joel Phillips. This project is supported by NSF grant 
DMR-9402020, AFOSR grant F49620-95-1-0311, and ARPA Contract 
N00174-93-C-0035.

\begin{figure}
 \caption {(a) The potential distribution due to a soliton located at 
the center ($i = 11$) of a $21 \times 1$ array. The circular symbol 
corresponds to the full capacitance matrix calculation and the square 
is for nearest neighbor approximation. The inset is a sketch of one section 
of our array. (b) The soliton potential distribution from the full 
capacitance matrix calculations (circles) compared with the predictions 
of Eq.~(\ref{eq:coulomb})(crosses).\label{fig:soliton}} 
\end{figure}

\begin{figure}
\caption{The soliton potential distributions in a $11 \times 11$ 2D array of 
quantum dots, using the full capacitance matrix (solid line mesh) and the 
nearest neighbor approximation (dashed line mesh). The soliton 
is located at the center of the array ($x = y = 0$). 
\label{fig:2d}}
\end{figure}

\begin{figure}
\caption{Current-voltage characteristics of a $21 \times 1$ quantum 
dot array using, the full capacitance matrix (solid line) and the nearest 
neighbor approximation (dashed line). The inset shows the I-V at a larger 
scale.\label{fig:iv}}
\end{figure}

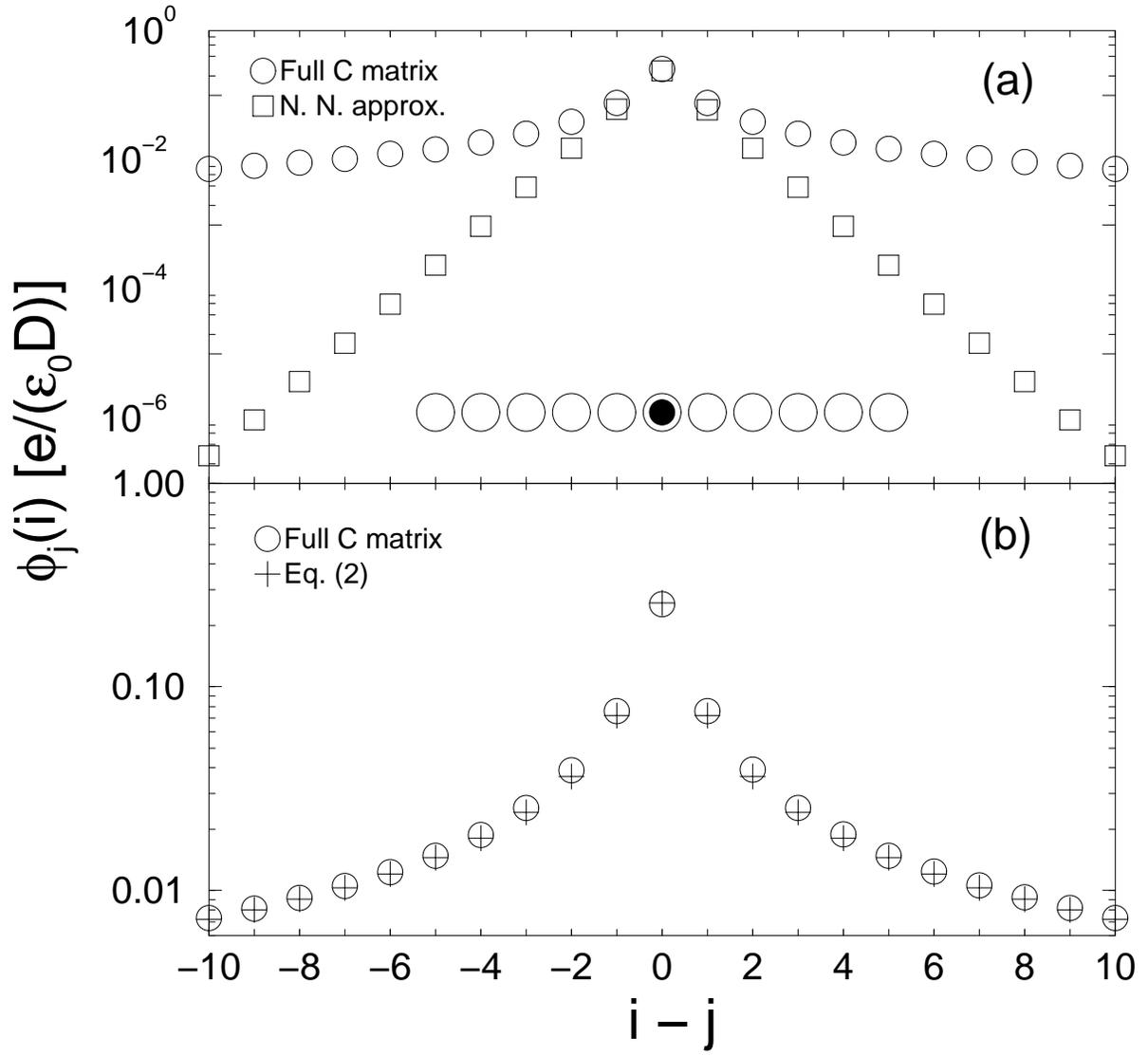
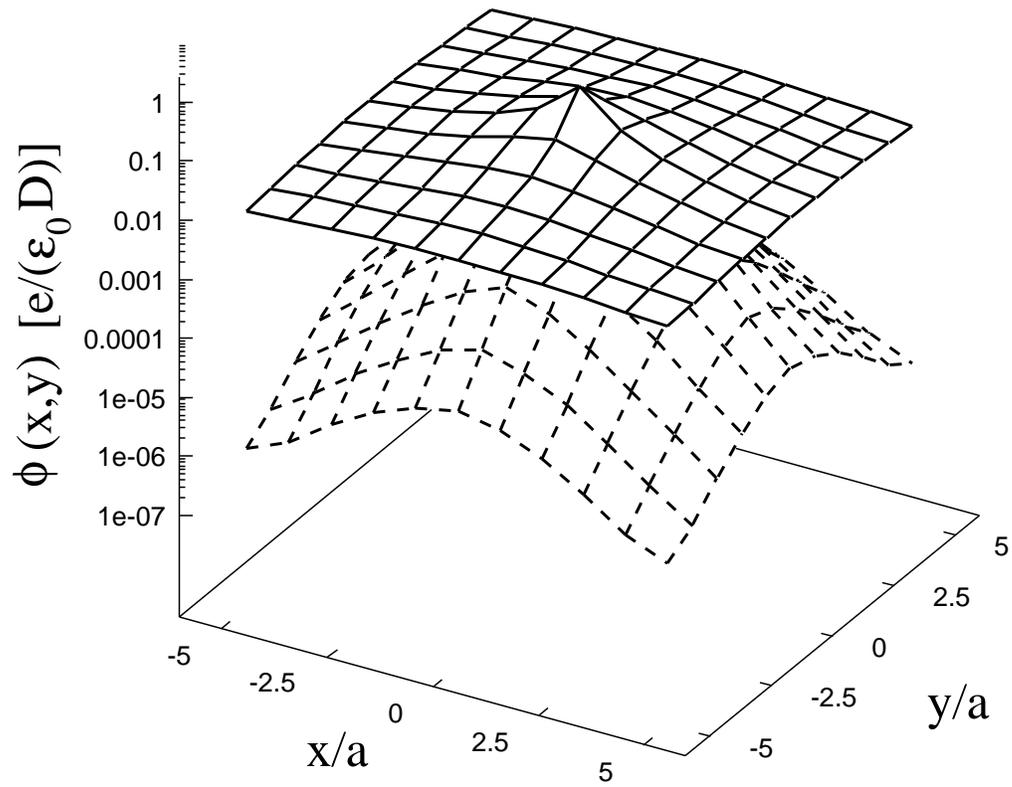
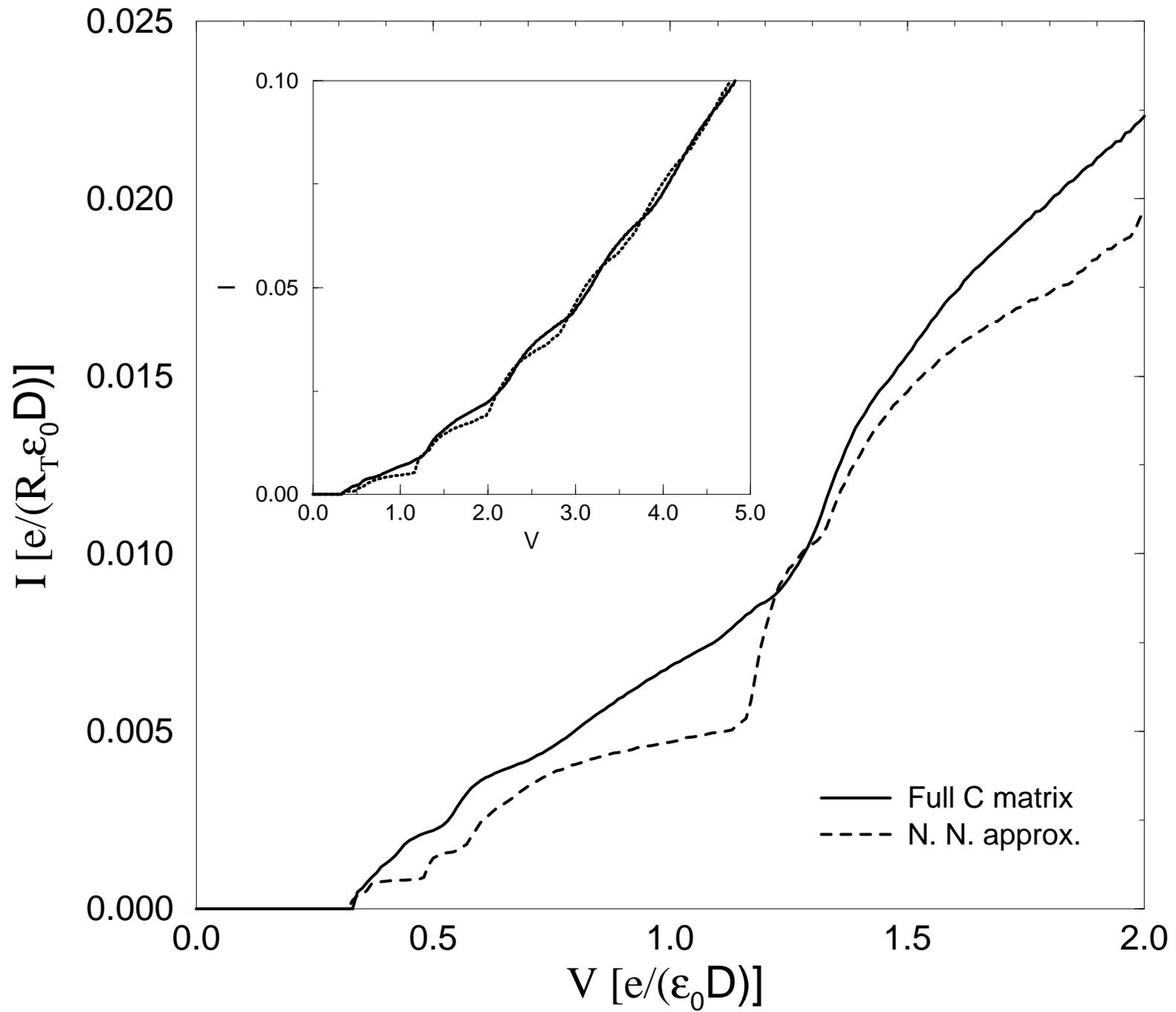
\end{document}